\documentclass[jcp,twocolumn,superscriptaddress,floatfix]{revtex4}
\usepackage{graphics,graphicx,dcolumn,bm,fleqn,epic,eepic,float}
\usepackage{amssymb,amsmath,multirow,rotate,color,float}
\usepackage[latin1]{inputenc}

\usepackage[dvips]{epsfig}

\begin{document}

\newcommand{\FIXME}[1]{\textcolor{red}\bf #1}

\title{Lattice-Boltzmann simulations of the drag force on a sphere approaching a superhydrophobic striped plane}
\author{Alexander L. Dubov}
\affiliation{A.N.~Frumkin Institute of Physical Chemistry and Electrochemistry, Russian Academy of Sciences, 31 Leninsky
Prospect, 119071 Moscow, Russia}

\affiliation{DWI - Leibniz Institute for Interactive Materials, RWTH Aachen, Forckenbeckstra\ss e 50, 52056 Aachen,
  Germany}

\author{Sebastian Schmieschek}
\affiliation{Department of Applied Physics, Eindhoven University of Technology, P.O. Box 513, 5600 MB Eindhoven, The Netherlands}

\author{Evgeny S. Asmolov}
\affiliation{A.N.~Frumkin Institute of Physical Chemistry and Electrochemistry, Russian Academy of Sciences, 31 Leninsky Prospect, 119071 Moscow, Russia}
\affiliation{Central Aero-Hydrodynamic Institute, 140180 Zhukovsky,
  Moscow region, Russia}
\affiliation{Institute of Mechanics, M.V. Lomonosov Moscow State
  University, 119991 Moscow, Russia}

\author{Jens Harting}
\affiliation{Department of Applied Physics, Eindhoven University of Technology, P.O. Box 513, 5600 MB Eindhoven, The Netherlands}
\affiliation{Institute for Computational Physics, University of Stuttgart,
Allmandring 3, 70569 Stuttgart, Germany}

\author{Olga I. Vinogradova}
\affiliation{A.N.~Frumkin Institute of Physical Chemistry and Electrochemistry, Russian Academy of Sciences, 31 Leninsky
Prospect, 119071 Moscow, Russia}

\affiliation{DWI - Leibniz Institute for Interactive Materials, RWTH Aachen, Forckenbeckstra\ss e 50, 52056 Aachen,
  Germany}

  \affiliation{Department of Physics, M.V.~Lomonosov Moscow State University, 119991 Moscow, Russia}


\begin{abstract}
By means of lattice-Boltzmann simulations the drag force on a sphere of
radius $R$ approaching a superhydrophobic striped wall has been investigated as
a function of arbitrary separation $h$. Superhydrophobic (perfect-slip vs.
no-slip) stripes are characterized by a texture period $L$ and a fraction of
the gas area $\phi$. For very large values of $h/R$ we recover the macroscopic
formulae for a sphere moving towards a hydrophilic no-slip plane. For $h/R
=O(1)$ and the drag force is smaller than predicted by classical
theories for hydrophilic no-slip surfaces, but larger than expected for a
sphere interacting with a uniform perfectly slipping wall. At a thinner gap, $h
\ll R$ the force reduction compared to a classical result becomes more
pronounced, and is maximized by increasing $\phi$. In the limit of very small separations our simulation
data are in quantitative agreement with an asymptotic equation, which relates a
correction to a force for superhydrophobic slip to texture parameters. In
addition, we examine the flow and pressure field and observe their  oscillatory
character in the transverse direction in the vicinity of the wall, which
reflects the influence of the heterogeneity and anisotropy of the striped
texture. Finally, we investigate the lateral force on the sphere, which is
detectable in case of very small separations and is
maximized by stripes with $\phi=0.5$.
\end{abstract}
\maketitle

\section{Introduction}

Superhydrophobic (Cassie) surfaces are able to trap air at the liquid-solid
interface, leading to remarkable wetting properties, such as a very large water
contact angles and low hysteresis~\cite{quere.d:2005}. They can also have an impact on the dynamics of the liquid.  For instance, the water drop slides or rolls with
amazingly large velocity, and a drop hitting such a material just bounces
off~\cite{tsai.p:2010}.  These macroscopic dynamic studies raised a question of
a remarkable drag reducing ability of superhydrophobic materials, which could
be extremely important for microfluidic lab-on-a-chip
systems~\cite{vinogradova.oi:2011,rothstein.jp:2010,vinogradova.oi:2012,mchale.g:2010}.
The flow of liquids near superhydrophobic surfaces is a subject that currently
attracts much experimental~\cite{choi.ch:2006,karatay.e:2013,joseph.p:2006},
simulation~\cite{priezjev.n:2011,zhou.j:2012,Schmieschek2012}, and
theoretical~\cite{ybert.c:2007,cottin.c:2004,feuillebois.f:2009,Asmolov:2012,cottin.c:2012,crowdy.d:2010}
research efforts.

Such a superlubrication potential could also dramatically modify a hydrodynamic
interaction between different surfaces. Therefore, it is attractive to consider
the hydrodynamic interaction of a hydrophilic sphere with a superhydrophobic
surface. Such a configuration is relevant for many surface forces
apparatus and atomic force microscope dynamic force
experiments~\cite{steinberger.a:2007,maali.a:2012}, it represents a typical
situation of phenomena of ``viscous adhesion'', coagulation, and more. However,
despite its importance for force experiments and numerous applications, the
quantitative understanding of the problem is still challenging.

The exact solution, valid for an arbitrary separation, for the drag force on a sphere moving
towards a flat wall is known only for a situation when both interacting surfaces are hydrophilic (i.e. characterized by hydrodynamic no-slip boundary conditions). This
was derived by Brenner~\cite{brenner61} and Maude~\cite%
{maude61} and is given by%
\begin{widetext}
\begin{eqnarray}
&&\!\!\!\!\!\!\!\!\!\!\!\ F_{BM}=-\frac{1}{3}F_{St}\sinh \xi
\left( \sum_{n=1}^{\infty }\frac{n(n+1)\left[
8e^{(2n+1)\xi }+2(2n+3)(2n-1)\right] }{(2n-1)(2n+3)[4\sinh ^{2}(n+\frac{1}{2}%
)\xi -(2n+1)^{2}\sinh ^{2}\xi ]}\right.   \notag \\
\label{eq:maudea} \\
&&\!\!\!\!\!\!\!\left. -\sum_{n=1}^{\infty }\frac{n(n+1)\left[
(2n+1)(2n-1)e^{2\xi }-(2n+1)(2n+3)e^{-2\xi }\right] }{(2n-1)(2n+3)[4\sinh
^{2}(n+\frac{1}{2})\xi -(2n+1)^{2}\sinh ^{2}\xi ]}\right) ,  \notag
\end{eqnarray}%
\end{widetext}
Here, $F_{St}$ is the Stokes drag on a sphere moving in an unbounded fluid,
\begin{equation}
F_{St}=6\pi \mu vR,
\end{equation}%
$\mu $ is the dynamic viscosity of the liquid, $v$ and $R$ are the velocity
and the radius of the sphere, $\cosh \xi =h/R$, $\xi <0$, and $h$ is the
distance between the apex of the sphere and the wall.

Let us remark that at large separations, $h\gg R,$ the two-term expansion of Eq. (\ref%
{eq:maudea}) gives~\cite{maude61}
\begin{equation}
F_{M} \simeq F_{St}\left( 1+\frac{9}{8}\frac{R}{h}\right) .  \label{firstorder}
\end{equation}%
Eq. (\ref{firstorder}) is surprisingly accurate even outside of the range of its formal applicability, and can be used in a very large interval of $h/R$.  However at small distances Eq.(\ref{firstorder}) gives $F_{M}/F_{St} \simeq 9R/(8h)$, and therefore deviates from the Taylor formula rigorously derived for the lubrication limit, $h/R\ll 1$:
\begin{equation}
F_{T} \simeq F_{St}\frac{R}{h}.  \label{Taylor}
\end{equation}%

Eqs.(\ref{eq:maudea}) and (\ref{Taylor}) have been verified for hydrophilic surfaces~\cite{Kunert2010,vinogradova:03,chan.dyc:1985,honing.cdf:2007}. However, their use in other situations  is beset with difficulties. There is a large literature describing attempts to answer questions of a validity of Eqs.(\ref{eq:maudea}) and (\ref{Taylor}) for hydrophobic and heterogeneous surfaces, and to provide a more general theory of hydrodynamic interactions. We mention below what we believe are the more relevant contributions, concentrating on the case of a hydrophilic sphere and hydrophobic or superhydrophobic flat wall.

Hydrophobic surfaces reduce drag, which is associated with a partial  slippage
of the fluid (characterized by a constant scalar slip
length)~\cite{vinogradova1999,boquet_farad_dis1999}. Although there is some
literature describing the motion of a hydrophilic sphere parallel to a
hydrophobic surface~\cite{davis:1994,luo2008,feuillebois2012}, information
about the motion towards a slippery wall is rather scarce.  Vinogradova
proposed a modification to the Taylor equation,
Eq.(\ref{Taylor})~\cite{vinogradova.oi:1995a,vinogradova:96}. She has argued
that it is convenient to describe a modification of a drag force in terms of a
correction for slippage:
\begin{equation}\label{fstar}
 f^{\ast}\simeq\frac{F_z}{F_T},
\end{equation}
and suggested general analytical expressions to relate $f^{\ast}$ and the slip length of interacting bodies. In case of an interaction of a hydrophilic sphere with a hydrophobic plane with a slip length $b$,  the theory predicts
 \begin{multline}\label{fb0}
    f^{\ast}\simeq\frac{F}{F_T}= \frac{1}{4} \left(1+\frac{3 h}{2 b} \left[ \left(1 + \frac{h}{4 b}\right)\times \right. \right. {} \\
   {} \left. \left. \times\ln\left(1+\frac{4 b}{h}\right)-1 \right] \right).
\end{multline}
In what follows $f^{\ast}$ can significantly decrease the hydrodynamic resistance force provided $h$ is of the order of $4b$ or smaller. In case of a perfect slip, $b=\infty$, Eq.(\ref{fb0}) predicts $f^{\ast}=1/4$. The
drag force however still remains inversely dependent on the gap as it is predicted by the Taylor theory. Later Lauga and Squires studied a sphere approaching a hydrophobic wall at very large separations, and obtained analytically a kind of small, of the order of $R/h$, correction to Eq.(\ref{firstorder})~\cite{lauga2005}.
We are unaware of any previous work that has addressed the question of hydrodynamic interaction with a hydrophobic plane at arbitrary separation and sphere radius. Note however that by symmetry reasons the problem of an interaction of a sphere with a perfectly slipping wall ($b=\infty$) is equivalent to that of the motion of
two equal spheres (separated by a twice as large distance) towards each other~\cite%
{happel1965,jeffrey1984}.

The flow past superhydrophobic surfaces is more complex. In this situation it
is advantageous to construct the tensorial effective slip boundary condition, which mimics
the actual one along the true heterogeneous
surface~\cite{Bazant2008,vinogradova.oi:2011,kamrin.k:2010,Schmieschek2012}. During the last few years several papers have been concerned with the interaction of a sphere with heterogeneous superhydrophobic surfaces. Very recently Asmolov et al. developed a theory  in the
limit $h\ll R$, $L\ll R$, where $L$ is a period of the texture~\cite{Asmolov:2011}. For the general case $h=O(L)$ the force should be found numerically.
However, for some limiting cases the asymptotic equations have been derived. When the gap is large compared to the texture lengthscale, $L\ll h\ll R$, the force reads~\cite{Asmolov:2011,lecoq.n:2004}
\begin{equation}
f^{\ast} \simeq 1-\frac{b_{\mathrm{eff}}^{\parallel }+b_{%
\mathrm{eff}}^{\perp }}{2h},  \label{asymp_large}
\end{equation}%
with $b_{\mathrm{eff}}^{\parallel }$ longitudinal and $b_\mathrm{eff}^{\perp }$ transverse slip lengths. In the opposite
limiting case of a thin gap, $h\ll \min \{b,L\}\ll R$, the correction for a superhydrophobic slip depends on the fraction of a gas/liquid area only~\cite{belyaev.av:2010b,Asmolov:2011}.%

The purpose of this lengthy introduction has been to show that despite its
importance for dynamic force and particle suspensions experiments, the
hydrodynamic interactions of a hydrophilic sphere with a superhydrophobic (and
even hydrophobic) plane remain poorly understood. A key difficulty is that
there is no general analytical or semi-analytical theory describing
hydrodynamic flows even on smooth hydrophobic surfaces, and on composite
surfaces analytical results (and in fact even numerical results) for the drag
force have only been obtained in simple lubrication geometries and specific
limits.  Therefore, many aspects of a hydrodynamic interaction of a sphere with
superhydrophobic surfaces have been given insufficient attention. The present
paper employs the lattice Boltzmann simulation to extend and generalize earlier
analysis.

Our paper is arranged as follows. In Sec.II we define our system, summarize earlier relevant hydrodynamic relationships and present some new theoretical estimates. Sec.III discusses our simulation method and justifies the choice of parameters. Results are discussed in Sec.IV and we conclude in Sec.V.

\section{Model and Theoretical Estimates}

We first present our model and basic theoretical relationship relevant for our analysis. Here, we mostly make use of the results presented earlier with some new interpretation.

\begin{figure}[tbp]
\begin{center}
\includegraphics[width=6cm]{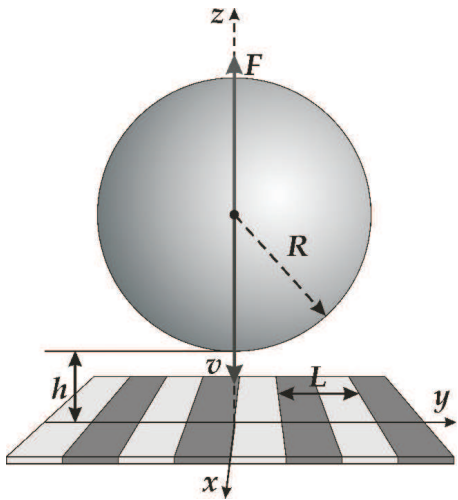}\\[0pt]
\end{center}
\caption{Sketch of the simulated system.
A sphere is approached towards a striped superhydrophobic wall.}
\label{fig_sketch}
\end{figure}

We consider a hydrodynamic interaction of a hydrophilic sphere with an idealized superhydrophobic surface in the Cassie state (sketched in Fig.~\ref{fig_sketch}),
where a liquid slab lies on top of the surface
roughness. To illustrate the approach, we focus here only on a periodically
striped surface. Such canonical textures are often employed in experiments and
are convenient to explore the basic physics of the system since the local
(scalar) slip length varies only in one direction and is piecewise constant, thus
allowing us
to highlight effects of anisotropy.

The liquid/gas interface is assumed to be flat with no meniscus
curvature, so that the modeled super-hydrophobic surface
appears as a perfectly smooth with a pattern of
boundary conditions. Similar assumptions have been made in most previous publications~\cite{ybert.c:2007,belyaev.av:2010b,priezjev.nv:2005,Asmolov:2012}. In this idealization,   we have neglected an additional mechanism for a dissipation connected with the meniscus curvature~\cite{harting.j:2008,sbragaglia.m:2007,ybert.c:2007}, which may have an influence
on a hydrodynamic force. Note however, that such a situation is not unrealistic, and has been achieved in many recent experiments~\cite{steinberger.a:2007,karatay.e:2013,haase.as:2013}.

Let $\mathbf{u}_{\tau }=\left( u_{x},u_{y},0\right) $ be the fluid velocity
along the wall, and $b\left( y\right) $ is the local slip length, which switches between two values, $b=\infty$ over gas/liquid regions and $b=0$ over solid/liquid
areas. This means that we set the shear-free boundary condition over the gas/liquid regions,
\begin{equation}\label{bcu2}
 \frac{\partial \mathbf{u}%
_{\tau }}{\partial z}=\mathbf{0},
\end{equation}
and the no-slip boundary condition at a solid/liquid interface,
\begin{equation}\label{bcu1}
  \mathbf{u}_{\tau }=\mathbf{0}.
\end{equation}
We remark that by assuming (\ref{bcu2}), the viscous
dissipations in the gas phase, which are expected to decrease the local slip length according to a ``gas cushion model''~\cite{vinogradova.oi:1995a}, have been neglected. Therefore our present results thus propose an upper bound for the local slip length at the gas areas. They could be generalized to the situation with
a finite (i.e. of the order of $L$ and smaller) slip length $b(y)$ on the gas/liquid interface, but we leave this generalization for future
work. By assuming no-slip, Eq.(\ref{bcu1}), at the solid area we neglect a
hydrophobic slippage, which is justified provided the nanometric slip length at
solid
areas is
small compared to a texture period~\cite{vinogradova.oi:2009,vinogradova:03,churaev.nv:1984,joly.l:2006}.

The fraction of gas/liquid area is given by $\phi$, and the fraction of
solid/liquid areas is then $1 - \phi$. When $\phi=0$ and $\phi=1$ the wall
becomes homogeneous.

In the limit of a thick channel ($h \gg L$) the eigenvalues of the slip-length tensor are given by~\cite{lauga.e:2003}
\begin{equation}\label{beff_ort_largeH_id}
  b_{\rm eff}^{\perp} \simeq \frac{L}{2 \pi} \ln\left[\sec\left(\displaystyle\frac{\pi \phi}{2 }\right)\right],\quad b_{\rm eff}^{\parallel}\simeq2 b_{\rm eff}^{\perp}.
\end{equation}
As proven in \cite{Belyaev2010a}, Eq.(\ref{beff_ort_largeH_id}) is also accurate and can safely be used in case of partial slip at the gas sectors, provided $b \gg L$.
The correction to Eq.~(\ref{Taylor}) is then
\begin{equation}
f^{\ast} \simeq 1 -\frac{3 L }{2h} \ln\left[\sec\left(\displaystyle\frac{\pi \phi}{2 }\right)\right].
\end{equation}%

In the limit of a thin channel ($h \ll L$) with one hydrophilic and one superhydrophobic wall the effective slip of a surface decorated by perfectly slipping stripes is~\cite{feuillebois.f:2009}

\begin{equation}\label{beff_ort_smallH_id}
  b_{\rm eff}^{\perp} \simeq h \frac{(1 - \phi)}{4 \phi},\quad b_{\rm eff}^{\parallel}\simeq 4 b_{\rm eff}^{\perp},
\end{equation}
and the correction for superhydrophobic slip takes the form~\cite{Asmolov:2011,belyaev.av:2010b}
\begin{equation}
f^{\ast}\simeq \frac{2(4-3\phi)}{8+9\phi-9\phi^{2}}. \label{b0}
\end{equation}

\section{Simulation}

In this section we present our simulation method and justify the choice of parameters.

For our computer experiment we chose a scheme based on
the lattice Boltzmann method which was successfully employed earlier in
comparable contexts~\cite{Kunert2010,Kunert2011,Schmieschek2012,Asmolov2013}.
Here, the lattice Boltzmann method serves as a Navier-Stokes solver for the
bulk fluid flow~\cite{bib:benzi-succi-vergassola}. Moving boundaries are employed
to describe the momentum exchange between the fluid and a suspended sphere
following the method summarized in Ref.~\cite{ladd_lattice-boltzmann_2001}. In addition,
an on-site slip boundary condition is used to describe the striped
substrate~\cite{Ahmed2009,Hecht2010}. A more detailed introduction to the simulation algorithm follows below.

The Boltzmann equation
\begin{equation}
\left[ \frac{\partial }{\partial t}+\mathbf{u}\cdot \nabla _{\mathbf{r}}%
\right] f(\mathbf{r,u},t)={\Omega},  \label{eq:boltzmann}
\end{equation}
expresses the dynamics of the single particle probability density
$f(\mathbf{r},\mathbf{u},t)$, where $\mathbf{r}$ is the position, $\mathbf{u}$
is the velocity, and $t$ is the time.
The left-hand side of Eq.~(\ref{eq:boltzmann}) models the propagation of
particles in phase space, while the right hand side accounts for particle
interactions by means of the collision operator $\Omega$.

By discretizing positions, velocities and time, a discrete variant of
Eq.~(\ref{eq:boltzmann}) can be obtained which is known as the lattice-Boltzmann
equation
\begin{equation}
\begin{array}{cc}
f_{k}(\mathbf{r}+\mathbf{c}_{k},t+\Delta t)-f_{k}(\mathbf{r},t)=\Omega _{k}, &
k=0,1,\dots ,B.%
\end{array}%
\end{equation}%
The lattice-Boltzmann equation describes the kinetics in discrete time-
($\Delta t$) and space-units ($ \Delta x$).  In the scope of this work we
employ the so-called D3Q19 lattice, referring to 3 dimensions and 19 discrete
velocity vectors $\mathbf{c}_{k};~k=\{0..18\}$ in direction of the $B=18$ nearest
neighbors of a cube as well as a zero velocity.
For convenience and without loss of generality, the scaling factors $\Delta x$
and $\Delta t$ are chosen to be of unity throughout the remainder of the text.

The use of the lattice Bhatnagar-Gross-Krook collision operator
\begin{equation}\label{eq:lbgk}
\Omega _{k}=-\frac{1}{\tau }(f_{k}(\mathbf{r},t)-f_{k}^{eq}(\mathbf{u}(%
\mathbf{r},t),\rho (\mathbf{r},t))),
\end{equation}
which assumes relaxation on a linear timescale $\tau$ towards a discretized
local Maxwell-Boltzmann distribution $f_{k}^{eq}$, is sufficient to recover a
second order accurate solution of the Navier-Stokes
equations~\cite{Bhatnagar1954}.  The kinematic viscosity $\nu =(2\tau -1)/6$ of
the fluid is related to the relaxation time scale. In this study the latter is
kept constant at $\tau =1$.  Macroscopic flow properties can be related to
stochastic moments of $f$, where the fluid density $\rho (\mathbf{r},t)=\rho
_{0}\sum_{k}f_{k}(\mathbf{r},t)$ and momentum
$\rho(\mathbf{r},t)\mathbf{u}(\mathbf{r},t)=\rho
_{0}\sum_{k}\mathbf{c}_{k}f_{k}(\mathbf{r},t)$ are of special interest ($\rho _{0}$ denotes a
reference density).

We use a simulation cell confined by two impermeable walls at $z=N_{z}$ and $z=0$.
At an upper wall we apply no-slip boundary conditions by means of
a simple mid-grid bounce back boundary condition. A bottom
wall is  modeled by patterns of a local slip length on a planar surface. More
specifically, we impose patches of zero and infinite slip-lengths by using (a
second order accurate) fixed velocity boundary condition as it was described
in~\cite{Hecht2010} and then used in several applications
in~\cite{Ahmed2009,Schmieschek2012,Asmolov2013}. The alternating perfect slip and no-slip stripes are aligned in \emph{y}-direction and
are of the widths $\phi L$ and $(1-\phi) L$, respectively. We vary $\phi$ with the
step size of 0.25 from 0 to 1. In $x$- and $y$-directions the simulation domain is limited by periodic
boundaries.

Following the approach of Ladd and
coworkers~\cite{ladd_lattice-boltzmann_2001}, the hydrophilic sphere is
implemented as a no-slip boundary moving with constant velocity $\bf v$. To
determine the momentum transferred to the fluid as the sphere moves, both the
center of mass velocity and the particle rotation are taken into account.
While the boundary moves on the fluid lattice, its discretization is constantly
adopted.
This causes fluctuations in the measured forces which we suppress by
averaging the measurement over approximately 1000
timesteps~\cite{Kunert2010,Kunert2011}.
The system resolution has to be balanced between an optimal approximation of
the model and computational cost.

\begin{figure}[tbp]
\begin{center}
\includegraphics[width=8cm]{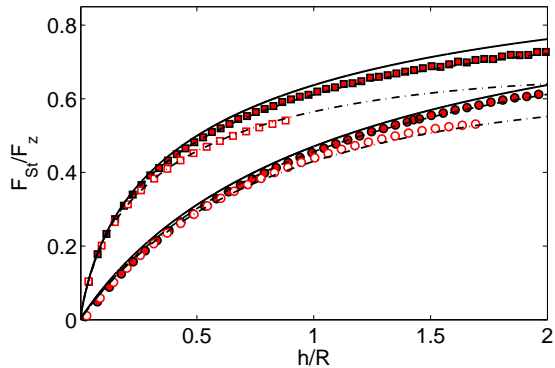}\\[0pt]
\end{center}
\caption{The drag force vs. separation curve obtained for two different heights
of the simulation cell, $N_z/R=8$ (open symbols) and $N_z/R=16$ (red symbols).
The upper data set corresponds to the interaction of a sphere with the no-slip wall.
The lower data set is obtained for the interaction with the perfectly slipping
wall.  Solid lines show the theoretical value calculated using
Eq.(\ref{eq:maudea}). Dashed-dotted lines are given by Eq.~(\ref{eq.2walls}). }
\label{fig.resolution}
\end{figure}

The cell height $N_{z}$ has to be sufficiently
large to allow to disregard the influence of the upper wall on the flow
field. However, it should be as small as possible to keep calculation times
within acceptable limits. To find the
most suitable $N_{z}$, we studied the influence of the upper surface on the results. When the sphere radius is small compared to the height of the box, the first-order correction to the drag due to the upper wall is inversely proportional to its separation from the sphere~\cite{happel1965}. We thus propose the following equation to fit the data:
\begin{equation}
F_z=F_{\rm theor} + F_{St} \frac{R}{N_z-2R-h},
\label{eq.2walls}
\end{equation}
where $F_{\rm theor}$ is the theoretical prediction for the drag in the absence of the upper wall.
Eq.~(\ref{eq.2walls}) allows us to estimate the effect
of the upper wall on the interaction for different $N_z$. We simulated the drag force acting on a sphere approaching the homogeneous bottom wall, and considered two
limiting cases. Namely, of a no-slip bottom wall, $\phi=0,$ and of a perfectly slipping wall, $\phi=1$, where exact theoretical solutions are known. We recall that in case of a no-slip wall the drag force is given
by Eq.(\ref{eq:maudea}), and that in case of perfectly slipping wall the drag force is equivalent to that for two approaching spheres at twice larger distance~\cite%
{happel1965,jeffrey1984}. We have plotted simulation results obtained with $N_z/R=8$ and $N_z/R=16$ in Fig.~\ref{fig.resolution}. Also included are the theoretical predictions and the theoretical values recalculated by using Eq.~(\ref{eq.2walls}). A general conclusion from this plot is that $N_z/R=16$ allows us to neglect the influence of the upper wall for $h/R<1$. For smaller cell height, $N_z/R\leq16$, or larger gaps, $h/R\geq1$, the influence of the upper boundary cannot be ignored. Note however that the data for $N_z/R=8$ are in good agreement with predictions of Eq.~(\ref{eq.2walls}).

Similar remarks concern the lateral size of the system, which should be sufficiently large to minimize
(without exceeding feasible computation times) the influence
of mirror spheres acting across the periodic boundaries. The periodic boundary conditions can be properly
accounted for by introducing an effective corrected radius, which difference from $R$ does not exceed
$2\%$ for $N_x/R=N_y/R=8$~\cite{Kunert2010,Kunert2011}. Finally, we note that the radius of the sphere should also be large enough to model analogy
with an actual sphere rather than a staircased construction. A further
requirement introduced by the theory is the radius to be significantly larger
than $L$. However, the width of a single
stripe has to be resolved by at least four lattice sites to avoid unacceptable
discretization errors~\cite{Ahmed2009,Kunert2010,Kunert2011}.

Taking into account above constraints and comparing different possible
configurations, we have determined a minimal set of simulation parameters given
by: $N_{x}=N_{y}=256$, $N_{z}=512$, $R=32$, $L=16$. Furthermore, selected runs
have been repeated at double resolution in order to approximate the
discretization errors introduced.

\section{Results and discussion}

In this Section we present our simulation results. More specifically, we
discuss simulation data obtained for normal and lateral forces in the system,
and also give some detailed analysis of a fluid velocity field and a pressure
distribution.

\subsection{Normal force}

\begin{figure}[tbp]
\begin{center}
\includegraphics[width=8cm]{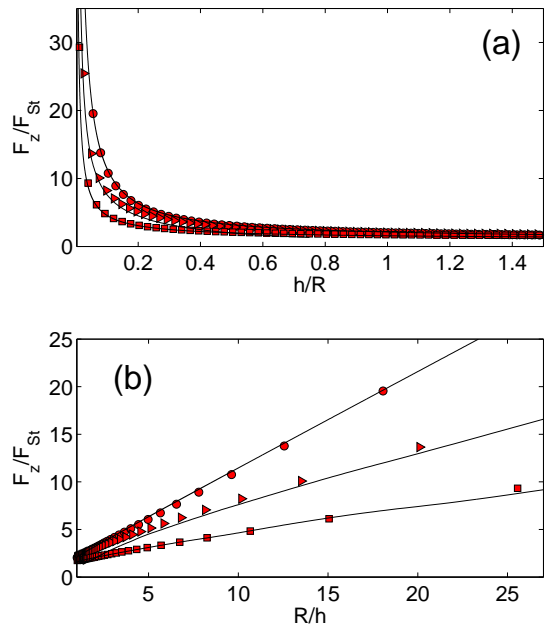}\\[0pt]
\end{center}
\caption{Hydrodynamic force on a sphere approaching a wall plotted in different coordinates. Symbols from top to bottom are the simulation data for $\phi=0, 0.5$ and $1$. Solid curves show theoretical predictions.}
\label{fig_fs1}
\end{figure}

Here we present the simulation results for the interaction of a sphere with striped
surfaces defined by different fractions of the gas phase.
In
order to assess the validity of the above lattice-Boltzmann approach, we first
measure the hydrodynamic force on a sphere approaching uniform non-slipping
($\phi=0$) and perfectly ($\phi=1$) smooth walls, where
theoretical solutions are known, i.e. verify Eq.(\ref{eq:maudea}) and results
of~\cite{happel1965,jeffrey1984}. Fig.~\ref{fig_fs1}(a) shows the simulation
results and theoretical curves for a normalized vertical force, $F_z/F_{St}$,
as a function of $h/R$. The quantitative agreement between the simulation and
theoretical results is excellent for all separations. This demonstrates the
accuracy of our simulations. It can be seen that in case of a perfectly
slipping wall the force at small distances is much smaller than predicted by
Eq.(\ref{eq:maudea}). To examine the short-distance region in more detail the
data from Fig.~\ref{fig_fs1}(a) are reproduced in Fig.~\ref{fig_fs1}(b) in
different coordinates. Fig.~\ref{fig_fs1}(b) is intended to indicate that
short-distance theoretical asymptotics for smooth surfaces is well reproduced
in simulations, and we emphasize that it is well seen that in
both cases when $h\ll R$ the force becomes inversely proportional to the gap as
predicted by Eqs.(\ref{Taylor}) and (\ref{fb0}). It is intuitively clear that
in case of patterned superhydrophobic surfaces the force should be confined
between these two limiting curves. The force curve simulated with $\phi=0.5$ is
included in Fig.~\ref{fig_fs1}. Note that here and in simulations below the
default location of the apex of the sphere is above the boundary between
no-slip and perfect-slip stripes (as shown in Fig.~\ref{fig_sketch}) unless
another configuration is specified.  The computed curve  indeed shows the drag
smaller than for a no-slip wall, but larger than for a perfectly slipping wall.
Fig.~\ref{fig_fs1} includes theoretical curves calculated within the
lubrication approach (i.e. justified provided $h \ll R$)~\cite{Asmolov:2011}. A
striking
result is that predictions of the lubrication-type theory are in a very good quantitative agreement with simulation results.

\begin{figure}[tbp]
\begin{center}
\includegraphics[width=8cm]{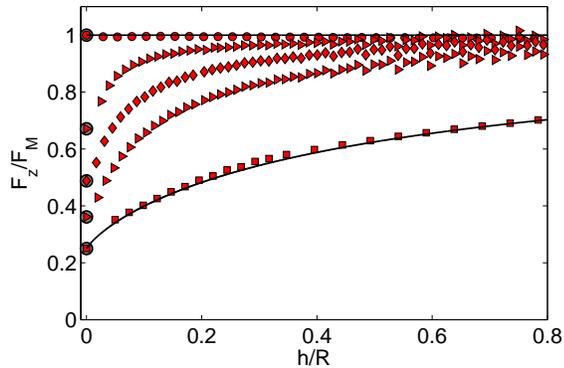}\\[0pt]
\end{center}
\caption{Correction for superhydrophobic slippage, $F_z/F_M$ vs. $h/R$ (red symbols). From top to bottom $\phi=0, 0.25, 0.5, 0.75,$ and $1$. Grey circles show the predictions of Eq.~(\ref{b0}). Solid curves plot the theoretical results for no-slip and perfectly slipping walls.}
\label{fig_fs2}
\end{figure}

We now investigate the effect of the gas fraction, $\phi$, on the drag force.
The measured data are presented in Fig.~\ref{fig_fs2}, which shows
$F_{z}/F_{M}$ as a function of $h/R$. The ratio $F_{z}/F_{M}$ denotes the
correction for superhydrophobic slippage to the force expected between two
hydrophilic surfaces. This is well seen in Fig.~\ref{fig_fs2}, which indicates
that $F_{z}/F_{M}=1$ at $\phi=0$. It can also be seen that the correction
factor decreases with $\phi$, i.e. the force becomes smaller than predicted by
Eq.(\ref{eq:maudea}). As expected, all curves are confined between two limiting
forces, obtained for $\phi=0$ and $\phi=1$. We should like to stress that the
correction for superhydrophobic slippage is long-range, and the deviations from
Eq.(\ref{eq:maudea}) are discernible even at $h/R = O(1)$, especially for large
$\phi$. At small $h/R$ this discrepancy becomes significant, and $F_{z}/F_{M}$
approaches its minimal value at a given $\phi$. In the lubrication limit, $R\gg
L\gg h$, these values can be evaluated with Eq.(\ref{b0}). Given the
approximations required to derive Eq.(\ref{b0}) we do not expect it to be
accurate for our system since the sphere radius in the simulation is not large
enough, $R/L=2$. However, calculations with  Eq.(\ref{b0}) shown in
Fig.~\ref{fig_fs2} coincide with the simulation. This surprising result
suggests that Eq.(\ref{b0}) is more general that it was assumed originally.

\subsection{Fluid velocity and pressure distribution }

\begin{figure}[tbp]
\begin{center}
\includegraphics[width=8cm]{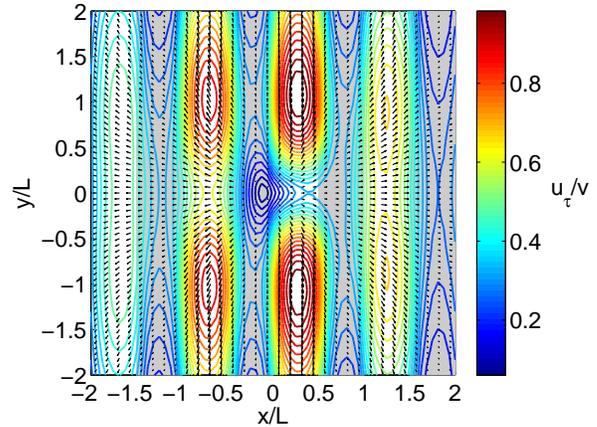}\\[0pt]
\end{center}
\caption{Vector and contour fields of the lateral velocity at the cross-section $z=L/8$ simulated for $h=3L/4$. No-slip stripes are shown by grey, and perfect-slip stripes by white color.}
\label{fig_vectorfield}
\end{figure}

Since our wall is a highly anisotropic striped surface it is instructive to study the velocity and excess (as compared to the ambient pressure at infinity) pressure, $p$, in the gap.

Fig.~\ref{fig_vectorfield} shows vector $\mathbf{u}_{\tau }$ and contour $|\mathbf{u}_{\tau }|$ fields measured for $h=3L/4$ in the
cross-section close to the wall, $z=L/8$. We see that the velocity is nonuniform
throughout the liquid. Note that there is a discernible asymmetry of the flow around the $x$-axis, which obviously reflects the location of the sphere apex above the edge of the stripes.
The velocity contour
lines are significantly elongated in the longitudinal direction, which
indicates that liquid preferably flows along the stripes. The velocity is very
small near the center, $x=y=0$, and at large distances from the axis,
$x^2+y^2\gg L^2$, where the gap becomes large. It has been earlier predicted
that for no-slip~\cite{chan.dyc:1985} and uniformly
slipping~\cite{horn.rg:2000} walls, where the flow is radially symmetric, and
at thin gaps, $h\ll R$, the maximum velocity is attained at distances from the
center of the order of $(Rh)^{1/2}$. It can be seen that the maximum velocity
in our case is also observed at distances of the order of $(Rh)^{1/2}$.
However, due to a strong anisotropy of the flow we also observe local velocity
maxima and minima over perfect-slip (white regions) and no-slip stripes (grey
regions), respectively.

\begin{figure*} [tbp]
\begin{center}
\includegraphics[width=11cm]{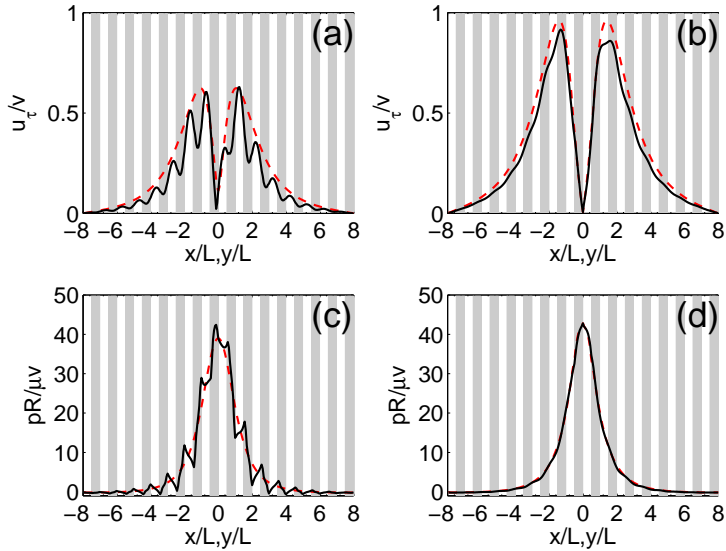}\\[0pt]
\end{center}
\caption{Distributions of lateral liquid velocity and pressure taken at the
cross-sections $z=L/8$ (a, c) and $z=L/2$ (b, d) and in longitudinal (dashed
curves) and transverse (solid curves) directions. Measurements are performed at
the fixed gap thickness $h=3L/4$. No-slip and perfect-slip regions are shown
by gray and white stripes.}
\label{fig_distribution}
\end{figure*}

To examine the significance of anisotropy of the texture in more details, we
have investigated the distributions of lateral velocity and the excess pressure in
the longitudinal and transverse directions.
Figs.~\ref{fig_distribution}(a) and (b) show the velocity in eigendirections
measured at two cross-sections. One is chosen to be in the vicinity of the wall $\left(
z_1=L/8\right) $, and another one is closer to the sphere $\left( z_2=L/2\right) $. It can be seen that at a small distance from
the striped wall (Fig.~\ref{fig_distribution}(a)), the longitudinal velocity
behaves qualitatively similar to what is expected for uniform
walls~\cite{chan.dyc:1985,horn.rg:2000}. More precisely, it is zero near the
origin of coordinates, has maxima at some distance from it, and decays at
larger distances. The transverse velocity shows irregular oscillatory behavior,
which is a reflection of the striped texture since the oscillations have a
period $L$. The oscillation maxima nearly coincide with the amplitude of the
longitudinal velocity and are detected at the middle of the perfect-slip
stripes. The minima are observed at the centers of no-slip regions, and here
the transverse velocity is much smaller than longitudinal. This result is similar to obtained earlier for weakly-slipping stripes~\cite{Asmolov2013}. At larger distance
from the striped wall (Figs.~\ref{fig_distribution}(b)), the flow is nearly
isotropic, and we do not observe any oscillation of the amplitude of the
transverse velocity. We also note that this cross-section is
characterized by a faster flow.

In Figs.~\ref{fig_distribution}(c) and (d) we plot the simulated
pressure at the same cross-sections. Perhaps the most important
conclusion from these plots is that despite an anisotropy of the flow, the
averaged over the texture period $L$ pressure  is
nearly isotropic.  Similar results have been already obtained numerically, but
only in the lubrication limit, $L,h \ll R$~\cite{Asmolov:2011}.  Another important point
to note is that pressure is the same for the two cross-sections, i.e. it does not
depend on $z$ being nearly constant across the gap, as in the lubrication theory.
Abrupt changes of the pressure in the transverse direction can be seen in
Fig.~\ref{fig_distribution}(c), which corresponds to the cross-section close to
the wall. In contrast to oscillations of a transverse velocity, the local
maxima and minima of pressure are located at the boundary between perfect-slip
and no-slip stripes. The pressure for flows across the stripes was predicted
to diverge at the wall, $z=0$, near the jump in $b(y)$: $p\propto r^{-1/2}$,
where $r\ll L$ is the distance from the border between
stripes~\cite{Asmolov:2012}. Our simulation results do confirm qualitatively
these theoretical predictions, but of course a quantitative agreement cannot be
obtained since we measure pressure at some finite distance from the wall, and
also because in simulations pressure is always finite.

\subsection{Lateral force}

\begin{figure*}[tbp]
\begin{center}
\includegraphics[width=18cm]{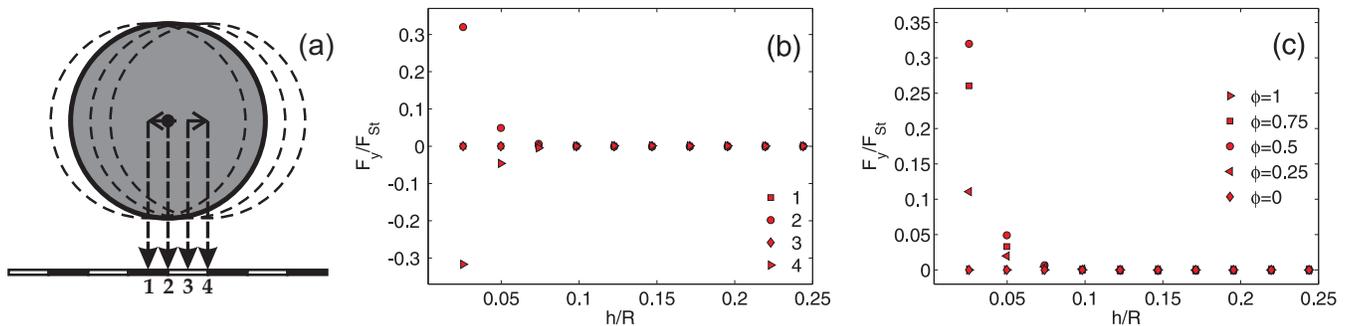}\\[0pt]
\end{center}
\caption{(a) Illustration of four possible sphere positions relative to the
texture. (b) Lateral forces at fixed $\phi=0.5$ simulated for different
locations of the sphere relative to stripes. (c) Lateral forces for a sphere in
position 2 simulated with different $\phi$.}
\label{fig_lateralforce}
\end{figure*}

Finally, we consider the lateral force on the sphere, which could be present in
addition to the normal drag force. In most real experiments the alignment of
the sphere and stripes is inconvenient or difficult, so the sphere could be in
different positions relative to the wall texture. In our simulations we explore
four representative cases as shown in Fig.~\ref{fig_lateralforce} (a). Two
configurations, 1 and 3, where the apex of the sphere is located above the
center of a no-slip or a perfect slip stripe, are symmetric. Therefore,  the
lateral force is expected to be absent. Since the asymmetry is maximized in two
other configurations, 2 and 4, where the sphere is above the border between
stripes, we consider them with the goal to maximize the lateral force. We
remark that our above analysis corresponds to a sphere in a position 2.

We show in  Fig.~\ref{fig_lateralforce} (b) the simulation
results for the lateral force obtained with $\phi=0.5$, which corresponds to
maximum transverse flow in a thin channel
situation~\cite{feuillebois.f:2010b,zhou.j:2012}. The simulation data demonstrates  that
the lateral force is measurable only when the sphere is very close to the wall,
$h/R \leq 0.05$, and that it is three orders of magnitude smaller than
the normal drag force. These results also clearly show the importance of a
location of a sphere relative to stripes in generating the lateral force. They
illustrate that the lateral force vanishes in positions 1 and 3, is positive
for  position 2, and negative for position 4. This indicates that in the
asymmetric position the lateral force pushes the sphere towards the center of
the perfect-slip stripes, where the friction is lower. We remark that recently
Pimponi et al. studied numerically the particle motion in the vicinity of the
striped wall and reported similar behavior of a lateral force of small
magnitude~\cite{pimponi.d:2013}.

The detailed comparison between the simulation results obtained with a different fraction of the gas/liquid area is
then shown in Fig.~\ref{fig_lateralforce}(c). We see that the maximal force is indeed obtained for the stripes with
equal area fractions, $\phi=0.5$, which correlates well with earlier theoretical predictions of transverse phenomena in thin channels~\cite{feuillebois.f:2010b}. The weaker lateral force is detected for $\phi=0.25$ and $0.75$, and it disappears for a homogeneous  surface, $\phi=0$ and $1$.

Finally, coming back to the normal force, a similar study of the role of the sphere location show that the changes in the normal force are detectable at $h \ll R$, but they never exceeds 1\% and can therefore safely be ignored. As a side note, one may remark that all our results have been obtained for $R = O(L)$ and could not be immediately applied to the situation of $R \ll L$, where the effect of a sphere location on the normal force might become significant. This case however is beyond the scope of present work and will be discussed elsewhere.

\section{Conclusion}

In conclusion, we have presented simulation data for a hydrodynamic interaction
of a sphere moving towards a superhydrophobic striped plane. We checked the
validity of our approach by reproducing the known theoretical predictions for
uniform no-slip and perfectly slipping walls. The simulation results show that
the drag force acting on a sphere approaching the striped wall is confined
between these two limiting solutions, and that the magnitude of the force
depends strongly on the fraction of gas/liquid areas. A quantitative agreement
with earlier predictions of a force in the limit of a thin gap has been
obtained. We have also examined the flow field and detected an oscillatory
character of its transverse component in the vicinity of the wall, which
reflects the influence of the heterogeneity and anisotropy of the striped
texture. Our analysis of pressure data led to a conclusion that despite an
anisotropy of the texture the \textbf{average} pressure in the gap remain surprisingly
isotropic. However, in the vicinity of the wall we observed abrupt jumps in
pressure in the transverse direction, which correlate well with earlier
predicted singularities at the border of no-slip and perfect-slip stripes.
Finally, we investigated the lateral force on the sphere, and found that it is
detectable in the case of the thin gap and also depends strongly on the
fraction of the gas areas.

\section{Acknowledgement}

This research was partly supported by the Russian Academy of Science (RAS)
through its priority program `Assembly and Investigation of Macromolecular
Structures of New Generations' (grant of O. I. Vinogradova), and by the Netherlands Organization for Scientific
Research (NWO/STW VIDI grant of J. Harting and NWO travel grant of E.S. Asmolov).
We acknowledge computing resources from the J\"ulich Supercomputing Center, the
Scientific Supercomputing Center Karlsruhe and SARA Amsterdam.


\end{document}